\documentclass[10pt, conference]{IEEEtran}
\IEEEoverridecommandlockouts

\usepackage{listings}
\DeclareFixedFont{\ttb}{T1}{txtt}{bx}{n}{8.5} % for bold
\DeclareFixedFont{\ttm}{T1}{txtt}{m}{n}{8.5}  % for normal

\usepackage{color}
\definecolor{deepblue}{rgb}{0,0,0.5}
\definecolor{deepred}{rgb}{0.6,0,0}
\definecolor{deepgreen}{rgb}{0,0.5,0}

\newcommand\pythonstyle{\lstset{
    language=Python,
    basicstyle=\ttm,
    aboveskip=10pt,
    belowskip=10pt,
    otherkeywords={self},             % Add keywords here
    keywordstyle=\ttb\color{deepblue},
    emph={MyClass,__init__},          % Custom highlighting
    emphstyle=\ttb\color{deepred},    % Custom highlighting style
    stringstyle=\color{deepgreen},
    numbers=none,
    frame=tb,                         % Any extra options here
    showstringspaces=false            % 
}}

\lstnewenvironment{python}[1][]
{
    \pythonstyle
    \lstset{#1}
}{}

\usepackage{color}
\usepackage{fancyvrb}
\usepackage{fvextra}

\DefineVerbatimEnvironment{Highlighting}{Verbatim}{commandchars=\\\{\}, breaklines=true}
\newenvironment{Shaded}{}{}

\newcommand{\DecValTok}[1]{\textcolor[rgb]{0.25,0.63,0.44}{{#1}}}

\newcommand{\NormalTok}[1]{{#1}}
\newcommand{\devito}{\href{https://github.com/devitocodes/devito}{Devito} }

\usepackage[unicode=true]{hyperref}
\hypersetup{breaklinks=true,
            bookmarks=true,
            pdfauthor={},
            pdftitle={Scaling through abstraction – high-performance vectorial wave simulations for seismic inversion with Devito.},
            colorlinks=true,
            citecolor=black,
            urlcolor=blue,
            linkcolor=black,
            pdfborder={0 0 0}}
\usepackage{cite}
\usepackage{amsmath,amssymb,amsfonts}
\usepackage{algorithmic}
\usepackage{graphicx}
\usepackage{textcomp}
\usepackage{xcolor}
\usepackage{longtable,booktabs}
\usepackage{caption}
\usepackage{subfig}
\def\BibTeX{{\rm B\kern-.05em{\sc i\kern-.025em b}\kern-.08em
    T\kern-.1667em\lower.7ex\hbox{E}\kern-.125emX}}

\begin{document}

\title{Scaling through abstractions – high-performance vectorial wave simulations for seismic inversion with Devito}

\author{
\IEEEauthorblockN{1\textsuperscript{st} Mathias Louboutin}
\IEEEauthorblockA{\textit{Georgia Institute of Technology}\\
Atlanta, GA \\
mlouboutin3@gatech.edu}
\and
\IEEEauthorblockN{2\textsuperscript{nd} Fabio Luporini}
\IEEEauthorblockA{\textit{Devito Codes}\\
London, UK \\
fabio@devitocodes.com}
\and
\IEEEauthorblockN{3\textsuperscript{rd} Rhodri Nelson}
\IEEEauthorblockA{\textit{Imperial College London}\\
London, UK \\
rhodri.nelson@imperial.ac.uk}
\and
\IEEEauthorblockN{4\textsuperscript{th} Philipp Witte}
\IEEEauthorblockA{\textit{Georgia Institute of Technology}\\
Atlanta, GA \\
pwitte3@gatech.edu}
\and
\IEEEauthorblockN{5\textsuperscript{th} George Bisbas}
\IEEEauthorblockA{\textit{Imperial College London}\\
London, UK \\
g.bisbas18@imperial.ac.uk}
\and
\IEEEauthorblockN{6\textsuperscript{th} Jan Thorbecke}
\IEEEauthorblockA{\textit{TU-Delft}\\
Delft, NL \\
J.W.Thorbecke@tudelft.nl}
\and
\IEEEauthorblockN{7\textsuperscript{th} Felix J. Herrmann}
\IEEEauthorblockA{\textit{Georgia Institute of Technology}\\
Atlanta, GA\\
felix.herrmann@gatech.edu}
\and
\IEEEauthorblockN{8\textsuperscript{th} Gerard J. Gorman}
\IEEEauthorblockA{\textit{Imperial College London}\\
London, UK \\
g.gorman@imperial.ac.uk}
}

% \and
% \and
% \and
% \and
% \and

\maketitle

\begin{abstract}\label{abstract}

\devito is an open-source
Python project based on domain-specific language and compiler
technology. Driven by the requirements of rapid HPC applications
development in exploration seismology, the language and compiler have
evolved significantly since inception. Sophisticated boundary
conditions, tensor contractions, sparse operations and features such as
staggered grids and sub-domains are all supported; operators of
essentially arbitrary complexity can be generated. To accommodate this
flexibility whilst ensuring performance, data dependency analysis is
utilized to schedule loops and detect computational-properties such as
parallelism. In this article, the generation and simulation of
MPI-parallel propagators (along with their adjoints) for the
pseudo-acoustic wave-equation in tilted transverse isotropic media and
the elastic wave-equation are presented. Simulations are carried out on
industry scale synthetic models in a HPC Cloud system and reach a
performance of 28TFLOP/s, hence demonstrating Devito's suitability for
production-grade seismic inversion problems.
\end{abstract}

\section{Introduction}\label{introduction}

Seismic imaging methods such Reverse Time Migration (RTM) and
Full-waveform inversion (FWI) rely on the numerical solution of an
underlying system of partial differential equations (PDEs), most
commonly some manifestation of the wave-equation. In the context of FWI,
the finite-difference (FDM) and the spectral-element (SEM) methods are
most frequently used to solve the wave-equation, with FDM methods
dominating within the seismic exploration community \cite{lyu2020}.
Various forms of the wave-equation and modelling strategies for FWI are
detailed in \cite{fichtner2011}.

Despite the theory of FWI dating back to the 1980s
\cite{tarantola}, among the first successful
expositions on real 3D data was presented in \cite{sirgue}. Other
studies utilizing FDM within the FWI workflow include
\cite{ratcliffe2011, petersson2013}. The aforementioned studies
approximate the underlying physics via the acoustic wave-equation;
higher fidelity models solving the non-isotropic elastic wave-equation
have been developed in, e.g.,
\cite{osti_1468379, osti_1561580, osti_1561581, sava1, sava2}. Owing to
the flexibility of the mathematical discretizations that can be
utilized, along with the ability to describe problems on complex meshes,
there has also been a great deal of interest in utilizing SEM to solve
inversion problems \cite{peter2011, krebsdg}. The recent study
\cite{trinh2019} presents an efficient SEM based inversion scheme using
a viscoelastic formulation of the wave-equation.

It is generally accepted that the PDE solver utilized within an
inversion workflow must satisfy the following criteria
\cite{virieuxmodelling}: - Efficient for multiple-source modelling - The
memory requirement of the modelling - The ability of a parallel algorithm
to use an increasing number of processors - Ability of the method to
process models of arbitrary levels of heterogeneity - Reduce the
nonlinearity of FWI - Feasibility of the extension of the modelling
approach to more realistic physical descriptions of the media.

It is with these specifications in mind that Devito, a symbolic domain
specific language (DSL) and compiler for the automatic generation of
finite-difference stencils, has been developed. Originally deigned to
accelerate research and development in exploration geophysics, the
high-level interface, previously described in detail in \cite{devito-api},
is built on top of \texttt{SymPy} \cite{sympy} and is inspired by the
underlying mathematical formulations and other DSLs such as FEniCS
\cite{fenics} and Firedrake \cite{firedrake}. This interface allows the
user to formulate wave-equations, and more generally time-dependent PDEs
in a simple and mathematically coherent way. The \devito compiler then
automatically generates finite-difference stencils from these mathematical
expressions. One of the main advantages of \devito over other
finite-difference DSLs is that generic expressions such as sparse
operations (i.e.~point source injection or localized measurements) are
fully supported and expressible in a high-level fashion. The second
component of \devito is its compiler (c.f \cite{devito-compiler}) that
generates highly optimized C code. The generated code is then compiled at
runtime for the hardware at hand.

Previous work focused on the DSL and compiler to highlight the potential
application and use cases of Devito. Here, we present a series of
extensions and applications to large-scale three-dimensional problem sizes
as encountered in exploration geophysics. These experiments are carried
out in Cloud-based HPC systems and include elastic forward modelling using
distributed-memory parallelism and imaging based on the tilted transverse
isotropic (TTI) wave-equation (\cite{virieux, thomsen1986, zhang-tti,
duveneck, louboutin2018segeow}). These proof of concepts highlight two
critical features: first, the ability of the symbolic interface and the
compiler to translate to large-scale adjoint-based inversion problems that
require massive compute (since thousands of PDEs are solved) as well as
large amounts of memory (since the adjoint state method requires the
forward model to be saved in memory). Secondly, through the elastic
modelling example, we demonstrate that \devito now fully supports and
automates vectorial and second order tensorial staggered-grid
finite-differences with the same high-level API previously presented for
scalar fields defined on cartesian grids.

This paper is organized as follows: first, we provide a brief overview of
Devito and its symbolic API and present the distributed memory
implementation that allows large-scale modelling and inversion by means of
domain decomposition. We then provide a brief comparison with a state of
the art hand-coded wave propagator to validate the performance previously
benchmarked with the roofline model (\cite{patterson, devito-compiler,
devito-api, louboutin2016ppf}). Before concluding, results from the
Cloud-based experiments discussed above are presented, highlighting the
vectorial and tensorial capabilities of Devito.

\section{Overview of Devito}\label{overview-of-devito}

Devito \cite{devito-api} provides a functional language built on top of
\texttt{SymPy} \cite{sympy} to symbolically express PDEs at a
mathematical level and implements automatic discretization with the
finite-difference method. The language is by design flexible enough to
express other types of non finite-difference operators, such as
interpolation and tensor contractions, that are inherent to
measurement-based inverse problems. Several additional features are
supported, among which are staggered grids, sub-domains, and stencils with
custom coefficients. The last major building block of a solid PDE solver
are the boundary conditions which for finite-difference methods are
notoriously diverse and often complicated. The system is, however,
sufficiently general to express them through a composition of core
mechanisms. For example, free surface and perfectly-matched layers
(PMLs) boundary conditions can be expressed as equations -- just like
any other PDE equations -- over a suitable sub-domain.

It is the job of the \devito compiler to translate the symbolic specification
into C code. The lowering of the input language to C consists of
several compilation passes, some of which introduce performance
optimizations that are the key to rapid code. Next to classic stencil
optimizations (e.g., cache blocking, alignment, SIMD and OpenMP
parallelism), \devito applies a series of FLOP-reducing transformations as
well as aggressive loop fusion. For a complete treatment, the interested
reader should refer to \cite{devito-compiler}.

\subsection{Symbolic language and
compilation}\label{symbolic-language-and-compilation}

In this section we illustrate the \devito language by demonstrating
an implementation of the acoustic wave-equation in isotropic media
\begin{equation}
\begin{cases}
 m \frac{d^2 u(t, x)}{dt^2} - \Delta u(t, x) = \delta(xs) q(t) \\
 u(0, .) = \frac{d u(t, x)}{dt}(0, .) = 0 \\
 d(t, .) = u(t, xr).
 \end{cases}
\label{acou}
\end{equation}
 The core of the Devito
symbolic API consists of three classes:

\begin{itemize}
\itemsep1pt\parskip0pt\parsep0pt
\item
  \texttt{Grid}, a representation of the discretized model.
\item
  \texttt{(Time)Function}, a representation of spatially (and time-)
  varying variables defined on a \texttt{Grid} object.
\item
  \texttt{Sparse(Time)Function} a representation of (time-varying) point
  objects on a \texttt{Grid} object, generally unaligned with respect to
  the grid points, hence called ``sparse''.
\end{itemize}

A \texttt{Grid} represents a discretized finite n-dimensional space and
is created as follows

\begin{python}[label=grid, caption=Grid creation]
from devito import Grid
grid = Grid(shape=(nx, ny, nz),
            extent=(ext_x, ext_y, ext_z),
            origin=(o_x, o_y, o_z))
\end{python}
\noindent
where \texttt{(nx,\phantom{\ }ny,\phantom{\ }nz)} are the number of grid
points in each direction,
\texttt{(ext\_x,\phantom{\ }ext\_y,\phantom{\ }ext\_z)} is the physical
extent of the domain in physical units (i.e \texttt{m}) and
\texttt{(o\_x,\phantom{\ }o\_y,\phantom{\ }o\_z)} is the origin of the
domain in the same physical units. The object \texttt{grid} contains all
the information related to the discretization such as the grid spacing.
We use \texttt{grid} to create the symbolic objects that will be used to
express the wave-equation. First, we define a spatially varying model
parameter \texttt{m} and a time-space varying field \texttt{u}

\begin{python}[label=function, caption=Function definition]
from devito import Function, TimeFunction
m = Function(name="m", grid=grid,
             space_order=so)
u = TimeFunction(name="u", grid=grid,
                 space_order=so,
                 time_order=to)
\end{python}
\noindent
where \texttt{so} is the order of the spatial discretization and \texttt{to}
the time discretization order used when generating the
finite-difference stencil. Next, we define point source objects (\texttt{src}) located at the physical
coordinates $x_r$, and the receiver (measurement) objects (\texttt{d})
located at the physical locations $x_r$

\begin{python}[label=sparse, caption=SparseFunction definition]
from devito import SparseTimeFunction
s = SparseTimeFunction(name="src",
                       grid=grid, npoint=1,
                       coordinates=x_s)
d = SparseTimeFunction(name="d", grid=grid,
                       npoint=1, nt=nt,
                       coordinates=x_r)
\end{python}

The source term is handled separately from the PDE as a point-wise
operation called injection, while measurement is handled via
interpolation. By default, \devito initializes all \texttt{Function} data
to 0, and thus automatically satisfies the zero Dirichlet condition at
\texttt{t=0}. The isotropic acoustic wave-equation can then be implemented
in \devito as follows

\begin{python}[label=WE, caption=Wave-equation symbolic definition]
from devito import solve, Eq, Operator
eq = m * u.dt2 - u.laplace
update = Eq(u.forward, solve(eq, u.forward))
src_eqns = s.inject(u.forward, expr=s*dt**2/m)
d_eqns = d.interpolate(u)
\end{python}

To trigger compilation one needs to pass the constructed equations to an
\texttt{Operator}.

\begin{python}[label=op, caption=Operator creation]
from devito import Operator
op = Operator(update + src_eqns + d_eqns)
\end{python}

The first compilation pass processes equations individually. The
equations are lowered to an enriched representation, while the
finite-difference constructs (e.g., derivatives) are translated into
actual arithmetic operations. Subsequently, data dependency analysis is
used to compute a performance-optimized topological ordering of the
input equations (e.g., to maximize the likelihood of loop fusion) and to
group them into so called ``clusters''. Basically, a cluster will
eventually be a loop nest in the generated code, and consecutive
clusters may share some outer loops. The ordered sequence of clusters
undergoes several optimization passes, including cache blocking and
FLOP-reducing transformations. It is then further lowered into an
abstract syntax tree, and it is on such representation that parallelism
is introduced (SIMD, shared-memory, MPI). Finally, all remaining
low-level aspects of code generation are handled, among which the most
relevant is data management (e.g., definition of variables,
transfers between host and device).

The output of the \devito compiler for the running example used in this
section is available at
\href{https://github.com/mloubout/SC20Paper/tree/master/codesamples}{CodeSample}
in \texttt{acou-so8.c}.

\subsection{Distributed-memory
parallelism}\label{distributed-memory-parallelism}

We here provide a succinct description of distributed-memory parallelism
in Devito; the interested reader should refer to the MPI tutorial at
\href{https://github.com/devitocodes/devito/blob/v4.2/examples/mpi/overview.ipynb}{mpi-notebook}
for thorough explanations and practical examples.

Devito implements distributed-memory parallelism on top of MPI. The design
is such that users can almost entirely abstract away from it and reuse
non-distributed code as is. Given any \devito code, just running it as

\begin{Shaded}
\begin{Highlighting}[]
\NormalTok{DEVITO_MPI=}\DecValTok{1} \NormalTok{mpirun -n X python ...}
\end{Highlighting}
\end{Shaded}

\noindent
triggers the generation of code with routines for halo exchanges. The
routines are scheduled at a suitable depth in the various loop nests
thanks to data dependency analysis. The following optimizations are
automatically applied:

\begin{itemize}
\itemsep1pt\parskip0pt\parsep0pt
\item
  redundant halo exchanges are detected and dropped;
\item
  computation/communication overlap, with prodding of the asynchronous
  progress engine by a designated thread through repeated calls to
  \texttt{MPI\_Test};
\item
  a halo exchange is placed as far as possible from where it is needed
  to maximize computation/communication overlap;
\item
  data packing and unpacking is threaded.
\end{itemize}

Domain decomposition occurs in Python upon creation of a \texttt{Grid}
object. Exploiting the MPI Cartesian topology abstraction, Devito
logically splits a grid based on the number of available MPI processes
(noting that users are given an ``escape hatch'' to override Devito's default
decomposition strategy). \texttt{Function} and \texttt{TimeFunction}
objects inherit the \texttt{Grid} decomposition. For
\texttt{SparseFunction} objects the approach is different. Since a
\texttt{SparseFunction} represents a sparse set of points, \devito looks at
the physical coordinates of each point and, based on the \texttt{Grid}
decomposition, schedules the logical ownership to an MPI rank. If a sparse
point lies along the boundary of two or more MPI ranks, then it is
duplicated in each of these ranks to be accessible by all neighboring
processes. Eventually, a duplicated point may be redundantly computed by
multiple processes, but any redundant increments will be discarded.

When accessing or manipulating data in a \devito code, users have the
illusion to be working with classic NumPy arrays, while underneath they
are actually distributed. All manner of NumPy indexing schemes (basic,
slicing, etc.) are supported. In the implementation, proper
global-to-local and local-to-global index conversion routines are used to
propagate a read/write access to the impacted subset of ranks. For
example, consider the array

\begin{Shaded}
\begin{Highlighting}[]
\NormalTok{A = [[ }\DecValTok{1}\NormalTok{,  }\DecValTok{2}\NormalTok{,  }\DecValTok{3}\NormalTok{,  }\DecValTok{4}\NormalTok{],}
     \NormalTok{[ }\DecValTok{5}\NormalTok{,  }\DecValTok{6}\NormalTok{,  }\DecValTok{7}\NormalTok{,  }\DecValTok{8}\NormalTok{],}
     \NormalTok{[ }\DecValTok{9}\NormalTok{, }\DecValTok{10}\NormalTok{, }\DecValTok{11}\NormalTok{, }\DecValTok{12}\NormalTok{],}
     \NormalTok{[}\DecValTok{13}\NormalTok{, }\DecValTok{14}\NormalTok{, }\DecValTok{15}\NormalTok{, }\DecValTok{16}\NormalTok{]])}
\end{Highlighting}
\end{Shaded}
\noindent
which is distributed across 4 ranks such that \texttt{rank\phantom{\ }0}
contains the elements reading
\texttt{1,\phantom{\ }2,\phantom{\ }5,\phantom{\ }6},
\texttt{rank\phantom{\ }1} the elements
\texttt{3,\phantom{\ }4,\phantom{\ }7,\phantom{\ }8},
\texttt{rank\phantom{\ }2} the elements
\texttt{9,\phantom{\ }10,\phantom{\ }13,\phantom{\ }14} and
\texttt{rank\phantom{\ }3} the elements
\texttt{11,\phantom{\ }12,\phantom{\ }15,\phantom{\ }16}. The slicing
operation \texttt{A{[}::-1,\phantom{\ }::-1{]}} will then return

\begin{Shaded}
\begin{Highlighting}[]
    \NormalTok{[[ }\DecValTok{16}\NormalTok{, }\DecValTok{15}\NormalTok{, }\DecValTok{14}\NormalTok{, }\DecValTok{13}\NormalTok{],}
     \NormalTok{[ }\DecValTok{12}\NormalTok{, }\DecValTok{11}\NormalTok{, }\DecValTok{10}\NormalTok{,  }\DecValTok{9}\NormalTok{],}
     \NormalTok{[  }\DecValTok{8}\NormalTok{,  }\DecValTok{7}\NormalTok{,  }\DecValTok{6}\NormalTok{,  }\DecValTok{5}\NormalTok{],}
     \NormalTok{[  }\DecValTok{4}\NormalTok{,  }\DecValTok{3}\NormalTok{,  }\DecValTok{2}\NormalTok{,  }\DecValTok{1}\NormalTok{]])}
\end{Highlighting}
\end{Shaded}
\noindent
such that now \texttt{rank\phantom{\ }0} contains the elements
\texttt{16,\phantom{\ }15,\phantom{\ }12,\phantom{\ }11} and so forth.

Finally, we remark that while providing abstractions for distributed
data manipulation, Devito
does not natively support any mechanisms for parallel I/O. However, the
distributed NumPy arrays along with the ability to seamlessly transfer any
desired slice of data between ranks provides a generic and flexible
infrastructure for the implementation of any form of parallel I/O
(e.g., see \cite{witte2018alf}).

\section{Industry-scale 3D seismic imaging in anisotropic
media}\label{industry-scale-3d-seismic-imaging-in-anisotropic-media}

One of the main applications of seismic finite-difference modelling in
exploration geophysics is reverse-time migration (RTM), a wave-equation based
seismic imaging technique. Real-world seismic imaging presents a number of
challenges that make applying this method to industry-scale problem sizes
difficult. Firstly, RTM requires an accurate representation of the underlying
physics via sophisticated wave-equations such as the tilted-transverse isotropic
(TTI) wave-equation, for which both forward and adjoint implementations must to
be provided. Secondly, wave-equations must be solved for a large number of
independent experiments, where each individual PDE solve is itself expensive in
terms of FLOPs and memory usage. For certain workloads, limited domain
decomposition, which balances the domain size and the number of independent
experiments, as well as checkpointing techniques must be adopted. In the following
sections, we describe an industry-scale seismic imaging problem that poses all
the aforementioned challenges, its implementation with Devito, and the results
of an experiment carried out on the Azure Cloud using a synthetic data set.

\subsection{Anisotropic wave-equation}\label{anisotropic-wave-equation}

In our seismic imaging case study, we use an anisotropic representation of the
physics called Tilted Transverse Isotropic (TTI) modelling \cite{thomsen1986}. This
representation for wave motion is one of the most widely used in exploration
geophysics since it captures the leading order kinematics and dynamics of
acoustic wave motion in highly heterogeneous elastic media where the medium
properties vary more rapidly in the direction perpendicular to sedimentary
strata \cite{alkhalifah2000, baysal1983, bubetti2012, bubetti2014,
bubesatti2016, chu2011, duveneck, fletcher, fowlertti2010,
louboutin2018segeow, whitmore1983, witte2016segpve, xu2014, zhang2005,
zhang2011, zhan2013}. The TTI wave-equation is an acoustic, low dimensional (4
parameters, 2 wavefields) simplification of the 21 parameter and 12 wavefields
tensorial equations of motions \cite{hooke}. This simplified representation is
parametrized by the Thomsen parameters $\epsilon(x), \delta(x)$ that relate to
the global (propagation over many wavelengths) difference in propagation speed
in the vertical and horizontal directions, and the tilt and azimuth angles
$\theta(x), \phi(x)$ that define the rotation of the vertical and horizontal
axes around the cartesian directions. However, unlike the scalar isotropic
acoustic wave-equation itself, the TTI wave-equation is extremely
computationally costly to solve and it is also not self-adjoint as shown in
\cite{louboutin2018segeow}.

The main complexity of the TTI wave-equation is that the rotation of the
symmetry axis of the physics leads to rotated second-order finite-difference
stencils. In order to ensure numerical stability, these rotated
finite-difference operators are designed to be self-adjoint (c.f.
\cite{zhang2011, duveneck}). For example, we define the rotated second order
derivative with respect to $x$ as:
\begin{equation}
\begin{aligned}
  G_{\bar{x}\bar{x}} &= D_{\bar{x}}^T D_{\bar{x}} \\
  D_{\bar{x}} &= \cos(\mathbf{\theta})\cos(\mathbf{\phi})\frac{\mathrm{d}}{\mathrm{d}x} + \cos(\mathbf{\theta})\sin(\mathbf{\phi})\frac{\mathrm{d}}{\mathrm{d}y} - \sin(\mathbf{\theta})\frac{\mathrm{d}}{\mathrm{d}z}.
\end{aligned}
\label{rot}
\end{equation}

We enable the simple expression of these complicated stencils in Devito
as finite-difference shortcuts such as \texttt{u.dx} where \texttt{u} is
a \texttt{Function}. Such shortcuts are enabled not only for the basic
types but for generic composite expressions, for example
\texttt{(u\phantom{\ }+\phantom{\ }v.dx).dy}. As a consequence, the
rotated derivative defined in~\ref{rot} is implemented with \devito in two
lines as:

\begin{python}[label=rotxpy, caption=Rotated finite-difference with symbolic shortcuts]
dx_u = cos(theta) * cos(phi) * u.dx +
       cos(theta) * sin(phi) * u.dy -
       sin(theta) * u.dz
dxx_u = (cos(theta) * cos(phi) * dx_u).dx.T +
        (cos(theta) * sin(phi) * dx_u).dy.T -
        (sin(theta) * dx_u).dz.T
\end{python}

Note that while the adjoint of the finite-difference stencil is enabled via the
standard Python \texttt{.T} shortcut, the expression needs to be reordered by hand
since the tilt and azymuth angles are spatially dependent and require to be
inside the second pass of first-order derivative. We can see from these simple
two lines that the rotated stencil involves all second-order derivatives
(\texttt{.dx.dx}, \texttt{.dy.dy} and \texttt{.dz.dz}) and all second-order
cross-derivatives (\texttt{dx.dy}, \texttt{.dx.dz} and \texttt{.dy.dz}) which
leads to a denser stencil support and higher computational complexity
(c.f. \cite{louboutin2016ppf}). For illustrative purposes, the complete
generated code for tti modelling with and without MPI is made available
at \href{https://github.com/mloubout/SC20Paper/tree/master/codesamples}{CodeSample}
in \texttt{tti-so8-unoptimized.c}, \texttt{tti-so8.c} and \texttt{tti-so8-mpi.c}.

Owing to the very high number of floating-point operations (FLOPs) needed per
grid point for the weighted rotated Laplacian, this anisotropic wave-equation
is extremely challenging to implement. As we show in Table~\ref{ttiFLOPs}, and
previously analysed in \cite{louboutin2016ppf}, the computational cost with
high-order finite-difference is in the order of thousands of FLOPs per grid
point without optimizations. The version without FLOP-reducing optimizations is
a direct translation of the discretized operators into stencil expressions (see
\texttt{tti-so8-unoptimized.c}). The version with optimizations employs
transformations such as common sub-expressions elimination, factorization, and
cross-iteration redundancy elimination -- the latter being key in removing
redundancies introduced by mixed derivatives. Implementing all of these
techniques manually is inherently difficult and laborious. Further, to obtain
the desired performance improvements it is necessary to orchestrate them with
aggressive loop fusion (for data locality), tiling (for data locality and
tensor temporaries), and potentially ad-hoc vectorization strategies (if
rotating registers are used). While an explanation of the optimization
strategies employed by \devito is beyond the scope of this paper (see
\cite{devito-compiler} for details), what is emphasized here is that users
can easily take full advantage of these optimizations without needed to
concern themselves with the details.

\begin{table}
\centering
\begin{tabular}{ccc}
\toprule\addlinespace
spatial order & w/o optimizations & w/ optimizations\tabularnewline
\midrule
4 & 501 & 95\tabularnewline
8 & 539 & 102\tabularnewline
12 & 1613 & 160\tabularnewline
16 & 5489 & 276\tabularnewline
\bottomrule
\end{tabular}
\caption{Per-grid-point FLOPs of the finite-difference TTI wave-equation stencil
with different spatial discretization orders.}\label{ttiFLOPs}
\end{table}

It is evident that developing an appropriate solver for the TTI wave-equation,
an endeavor involving complicated physics, mathematics, and engineering, is
exceptionally time-consuming and can lead to thousands of lines of code even for
a single choice of discretization. Verification of the results is no less
complicated, any minor error is effectively untrackable and any change to the
finite-difference scheme or to the time-stepping algorithm is difficult to
achieve without substantial re-coding. Another complication stems from the fact
that practitioners of seismic inversion are often geoscientists, not computer
scientists/programmers. Low level implementations from non-specialists can
often lead to poorly performing code. However, if research codes are passed to
specialists in the domain of low level code optimization they often lack the
necessary geophysical domain knowledge, resulting in code that may lack a key
feature required by the geoscientist. Neither situation is conducive to
addressing the complexities that come with implementing codes based on the
latest geophysical insights in tandem with those from high-performance
computing. With \devito on the other hand, both the forward and adjoint
equations can be implemented in a few lines of Python code as illustrated
with the rotated operator in Listing~\ref{rotxpy}. The low level optimization element
of the development is then taken care of under the hood by the Devito
compiler.

The simulation of wave motion is only one aspect of solving problems in
seismology. During wave-equation based imaging, it is also required to compute
sensitivities (gradient) with respect to the quantities of interest. This
requirement imposes additional constraints on the design and implementation of
model codes as outlined in \cite{virieux}. Along with several factors, such as
fast setup time, we focused on correct and testable implementations for the
adjoint wave-equation and the gradient (action of the adjoint Jacobian)
\cite{louboutin2018segeow, louboutin2020THmfi}; exactness being a mandatory
requirement of gradient based iterative optimization algorithms.

\subsection{3D Imaging example on Azure}\label{d-imaging-example-on-azure}

We now demonstrate the scalability of \devito to real-world applications by
imaging an industry-scale three-dimensional TTI subsurface model. This
imaging was carried out in the Cloud on Azure and takes advantage of
recent work to port conventional cluster code to the Cloud using a
serverless approach. The serverless implementation is detailed in
\cite{witte2019TPDedas, witte2019SEGedw} where the steps to run
computationally and financially efficient HPC workloads in the Cloud are
described. This imaging project, in collaboration with Azure, demonstrates
the scalability and robustness of \devito to large scale wave-equation
based inverse problems in combination with a cost-effective serverless
implementation of seismic imaging in the Cloud. In this example, we imaged
a synthetic three-dimensional anisotropic subsurface model that mimics a
realistic industry size problem with a realistic representation of the
physics (TTI). The physical size of the problem is
\texttt{10kmx10kmx2.8km} discretized on a \texttt{12.5m} grid with
absorbing layers of width 40 grid points on each side leading to
\texttt{881x881x371} computational grid points ($\approx300$ Million grid
points). The final image is the sum of 1500 single-source images: 100
single-source images were computed in parallel on the 200 nodes available
using two nodes per source experiment.
\vskip 0.1in
\noindent
\textbf{\emph{Computational performance}}
\vskip 0.1in
We briefly describe the computational setup and the performance achieved
for this anisotropic imaging problem. Due to time constraints, and because
the resources we were given access to for this proof of concept with Azure
were somewhat limited, we did not have access to Infiniband-enabled
virtual machines (VM). This experiment was carried out on
\texttt{Standard\_E64\_v3} and \texttt{Standard\_E64s\_v3} nodes which,
while not HPC VM nodes, are memory optimized thus allowing to the
wavefield to be stored in memory for imaging (TTI adjoint state gradient
\cite{virieux, louboutin2018segeow}). These VMs are Intel® Broadwell
E5-2673 v4 2.3GH that are dual socket, 32 physical cores (with
hyperthreading enabled) and 432Gb of DRAM. The overall inversion involved
computing the image for 1500 source positions, i.e.~solving 1500 forward
and 1500 adjoint TTI wave-equations. A single image required, in single
precision, 600Gb of memory. Two VMs were used per source and MPI set with
one rank per socket (4 MPI ranks per source) and 100 sources were imaged
in parallel. The performance achieved was as follows:

\begin{itemize}
\itemsep1pt\parskip0pt\parsep0pt
\item
  140 GFLOP/s per VM;
\item
  280 GFLOP/s per source;
\item
  28 TFLOP/s for all 100 running sources;
\item
  110min runtime per source (forward + adjoint + image computation).
\end{itemize}

We comment that if more resources were available, and because the imaging
problem is embarrassingly parallel over sources and can scale arbitrarily, the
imaging of all of the 1500 sources in parallel could have been attempted, which
theoretically leads to a performance of 0.4PFLOP/s.
\vskip 0.1in
\noindent
\textbf{\emph{How performance was measured}}
\vskip 0.1in
The execution time is computed through Python-level timers prefixed by an
MPI barrier. The floating-point operations are counted once all of the
symbolic FLOP-reducing transformations have been performed during
compilation. \devito uses an in-house estimate of cost, rather than
\texttt{SymPy}'s estimate, to take care of some low-level intricacies. For
example, Devito's estimate ignores the cost of integer arithmetic used for
offset indexing into multi-dimensional arrays. To calculate the total
number of FLOPs performed, \devito multiplies the floating-point operations
calculated at compile time by the size of the iteration space, and it does
that at the granularity of individual expressions. Thanks to aggressive
code motion, the amount of innermost-loop-invariant sub-expressions in an
\texttt{Operator} is typically negligible and therefore the \devito estimate
does not suffer from this issue, or at least not, to the best of our
knowledge, in a tangible way. The Devito-reported GFLOP/s were also
checked against those produced by Intel Advisor on several single-node
experiments: the differences -- typically \devito underestimating the
achieved performance -- were always at most in the order of units, and
therefore negligible.

\begin{figure*}
\centering
\captionsetup[subfigure]{labelformat=empty}
\subfloat[]{\includegraphics[width=0.500\hsize]{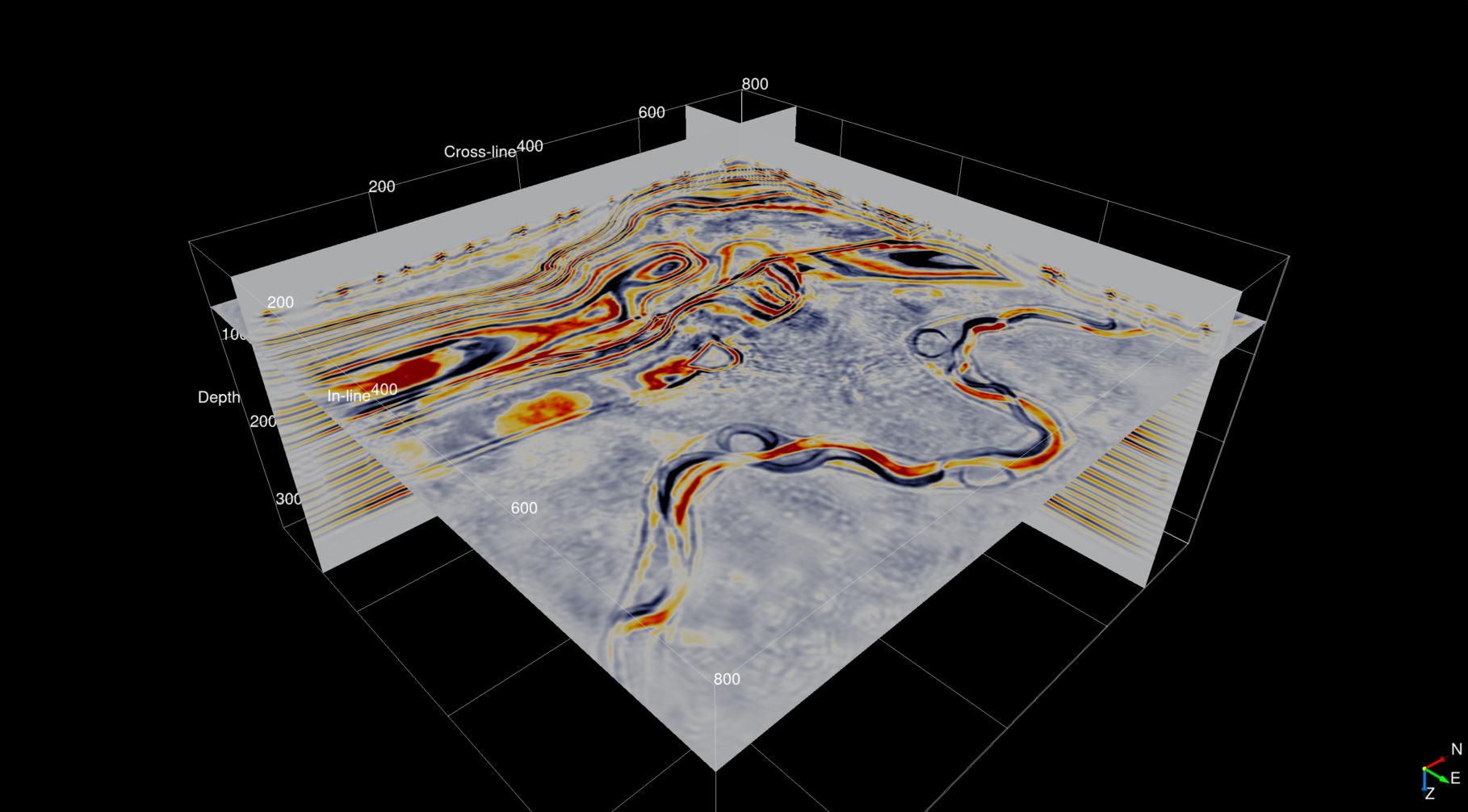}}
\subfloat[]{\includegraphics[width=0.500\hsize]{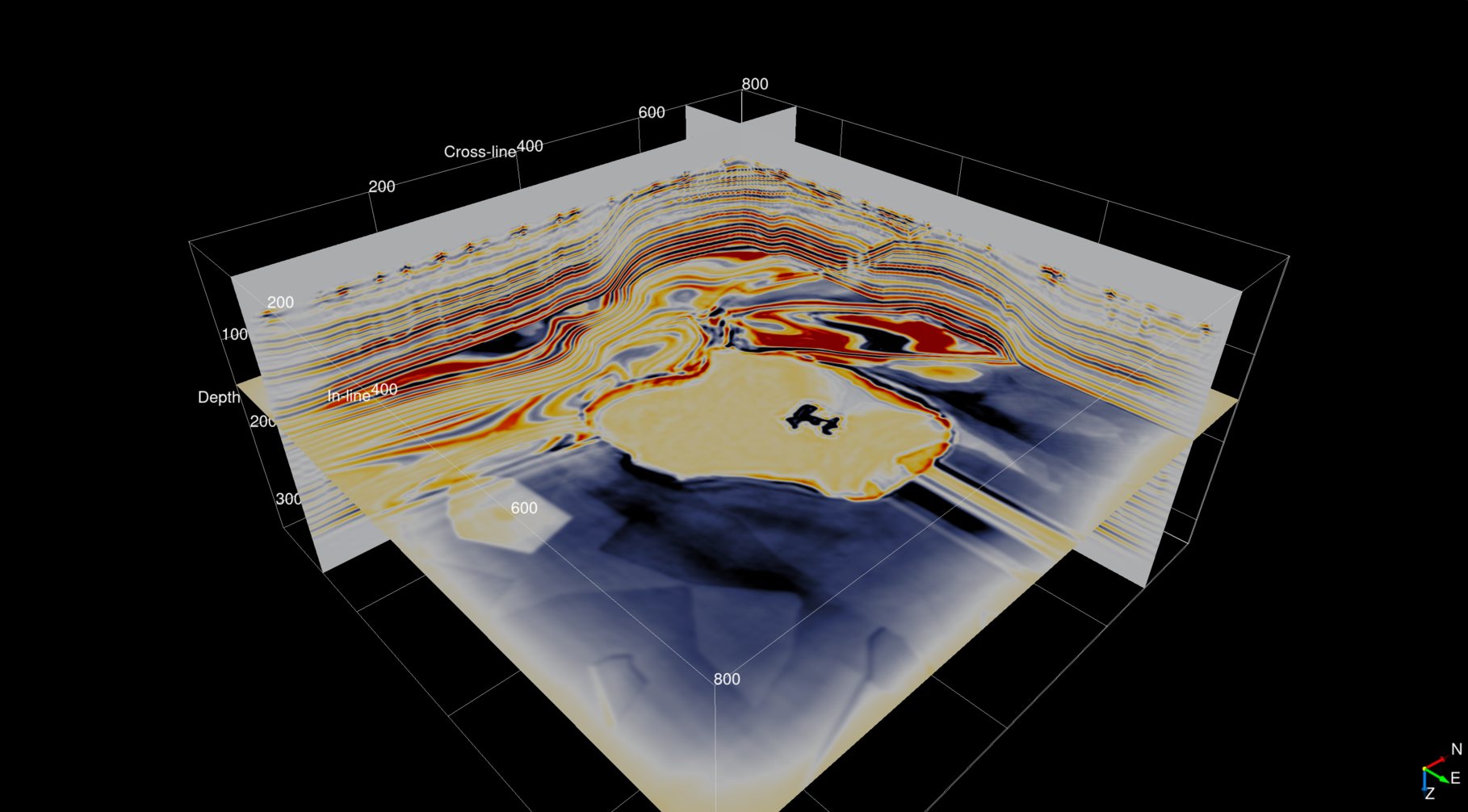}}
\\
\subfloat[]{\includegraphics[width=0.500\hsize]{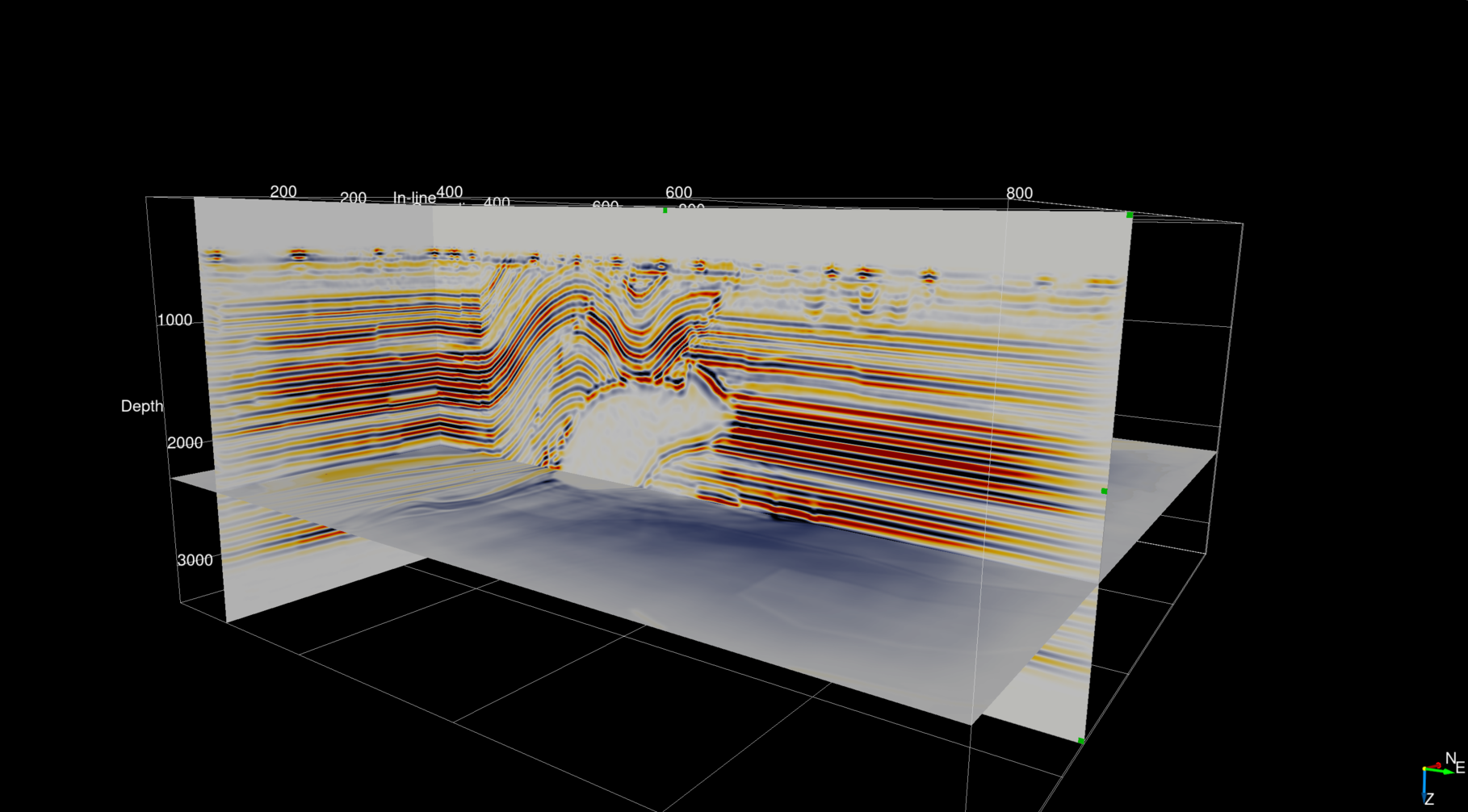}}
\subfloat[]{\includegraphics[width=0.500\hsize]{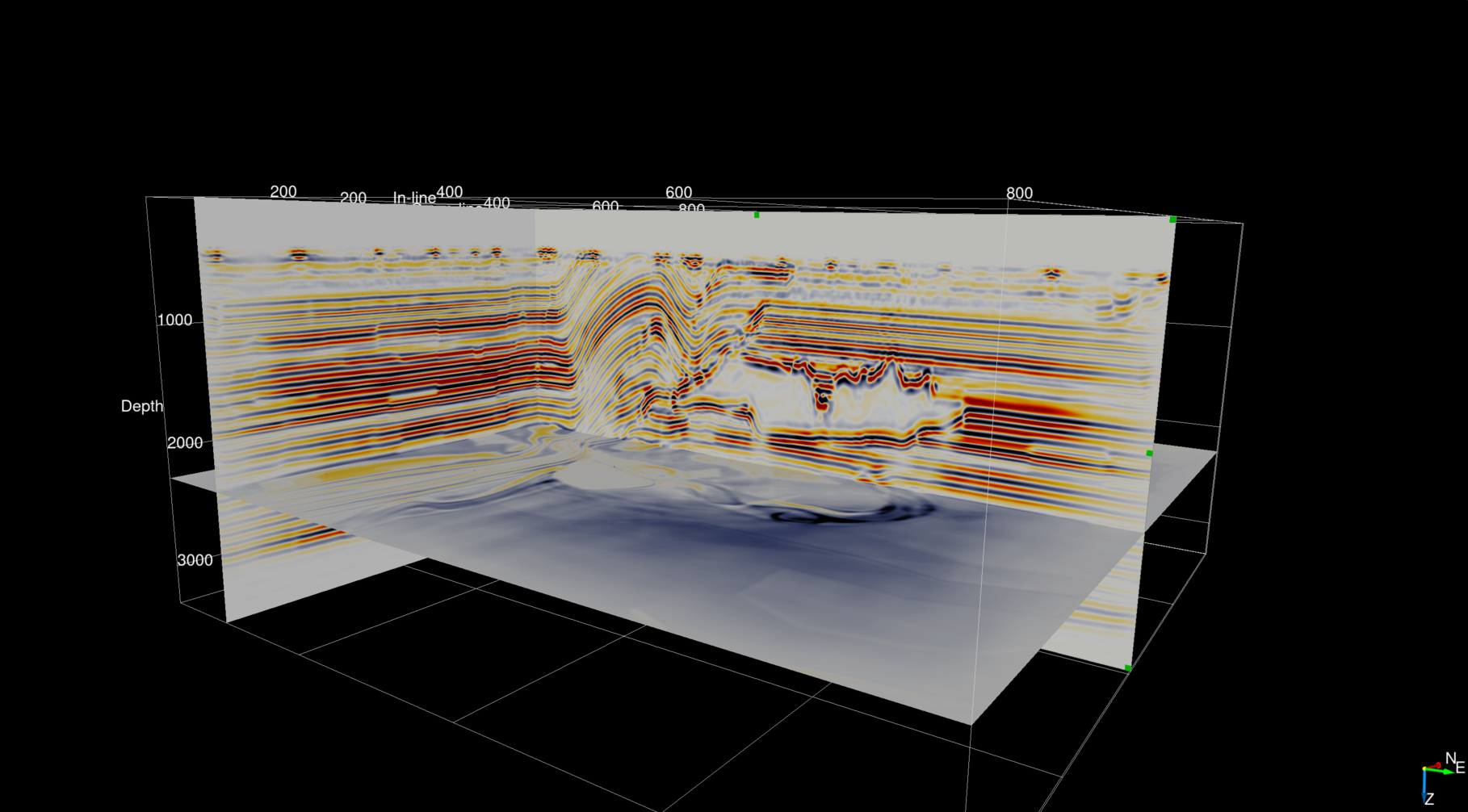}}
\caption{3D TTI imaging on a custom made model.}\label{OverTTI}
\end{figure*}

\vskip 0.1in
\noindent
\textbf{\emph{Imaging result}}
\vskip 0.1in
The subsurface velocity model used in this study is an artificial anisotropic
model that is designed and built combining two broadly known and used
open-source SEG/EAGE acoustic velocity models that each come with realistic
geophysical imaging challenges such as sub-salt imaging. The anisotropy
parameters are derived from a smoothed version of the velocity while the tilt
angles were derived from a combination of the smooth velocity models and
vertical and horizontal edge detection. The final seismic image of the
subsurface model is displayed in Figure~\ref{OverTTI} and highlights the fact
that 3D seismic imaging based on a serverless approach and automatic code
generation is feasible and provides good results.

\cite{witte2019TPDedas} describes the serverless implementation of seismic
inverse problems in detail, including iterative algorithms for least-square
minimization problems (LSRTM). The 3D anisotropic imaging results were presented
as part of a keynote presentation at the EAGE HPC workshop in October 2019
\cite{herrmann2019EAGEHPCaii} and at the Rice O\&G HPC workshop
\cite{witte2019RHPCssi} in which the focus was on the serverless implementation
of seismic inversion algorithms in the Cloud. This work illustrates the
flexibility and portability of Devito: we were able to easily port a code
developed and tested on local hardware to the Cloud, with only minor
adjustments. Further, note that this experiment included the porting of
MPI-based code for domain decomposition developed on desktop computers to the
Cloud. Our experiments are reproducible using the instructions in a public
repository \href{https://github.com/slimgroup/Azure2019/tree/v1.0}{AzureTTI}, which
contains, among the other things, the Dockerfiles and Azure
\href{https://batch-shipyard.readthedocs.io}{batch-shipyard} setup. This example
can also be easily run on a traditional HPC cluster environment using, for example, 
\href{https://github.com/slimgroup/JUDI.jl}{JUDI}~\cite{witte2018alf} or Dask~\cite{dask}
for parallelization over sources.

\section{Elastic modelling}\label{elastic-modelling}

While the subsurface image obtained in section \ref{d-imaging-example-on-azure}
utilized anisotropic propagators capable of mimicking intricate physics, in order
to model both the wave kinematics and amplitudes correctly, elastic
propagators are required. These propagators are, for example, extremely
important in global seismology since shear surface waves (which are
ignored in acoustic models) are the most hazardous. In this section, we
exploit the tensor algebra language introduced in \devito v4.0 to express
an elastic model with compact and elegant notation.

The isotropic elastic wave-equation, parametrized by the so-called Lamé parameters
$\lambda, \mu$ and the density $\rho$ reads:
\begin{equation}
\begin{aligned}
&\frac{1}{\rho}\frac{dv}{dt} = \nabla . \tau \\
&\frac{d \tau}{dt} = \lambda \mathrm{tr}(\nabla v) \mathbf{I}  + \mu (\nabla v + (\nabla v)^T)
\end{aligned}
\label{elas1}
\end{equation}
\noindent
 where $v$ is a vector valued function with one component per cartesian
direction:
\begin{equation}
\begin{split}
v =  \begin{bmatrix} v_x(t, x, y) \\ v_y(t, x, y)) \end{bmatrix}
\end{split}
\label{partvel}
\end{equation}
 and the stress $\tau$ is a symmetric second-order tensor-valued
function:
\begin{equation}
\begin{aligned}
    \tau = \begin{bmatrix}\tau_{xx}(t, x, y) & \tau_{xy}(t, x, y)\\\tau_{xy}(t, x, y) & \tau_{yy}(t, x, y)\end{bmatrix}.
\end{aligned}
\label{stress}
\end{equation}
 The discretization of \eqref{elas1} and \eqref{stress} requires five
equations in two dimensions (two equations for the particle velocity and
three for the stress) and nine equations in three dimensions (three for the particle velocity
and six for the stress). However, the mathematical
definition only require two coupled vector/tensor-valued equations for
any number of dimensions.

\subsection{Tensor algebra language}\label{tensor-algebra-language}

We have augmented the \devito language with tensorial objects to enable
straightforward and mathematically rigorous definitions of
high-dimensional PDEs, such as the elastic wave-equation defined in
\eqref{elas1}. This implementation was inspired by \cite{ufl}, a
functional language for finite element methods.

The extended \devito language introduces two new types, 
\texttt{VectorFunction}/\texttt{VectorTimeFunction} for vectorial objects such as the particle velocity, and
\texttt{TensorFunction}/\texttt{TensorTimeFunction} for second-order
tensor objects (matrices) such as the stress. These new objects are
constructed in the same manner as scalar \texttt{Function} objects. They
also automatically implement staggered grid and staggered
finite-differences with the possibility of half-node averaging. Each
component of a tensorial object -- a (scalar) \devito \texttt{Function} --
is accessible via conventional vector notation (e.g.
\texttt{v{[}0{]},\phantom{\ }t{[}0,1{]}}).

With this extended language, the elastic wave-equation defined in
\eqref{elas1} and \eqref{stress} can be expressed in only four lines
of code:

\begin{python}[label=elastic, caption=Vectorial elastic wave-equation with Devito]
from devito import VectorTimeFunction,
                   TensorTimeFunction
v = VectorTimeFunction(name="v", grid=grid,
                       space_order=so,
                       time_order=1)
tau = TensorTimeFunction(name="t", grid=grid,
                         space_order=so,
                         time_order=1)
u_v = Eq(v.forward,
         damp * (v + s/rho*div(tau)))
u_t = Eq(tau.forward,
         damp * (tau + s * (l * diag(div(v.forward)) + mu * (grad(v.forward) + grad(v.forward).T))))
\end{python}

\begin{figure*}
\centering
\includegraphics[width=1.000\hsize]{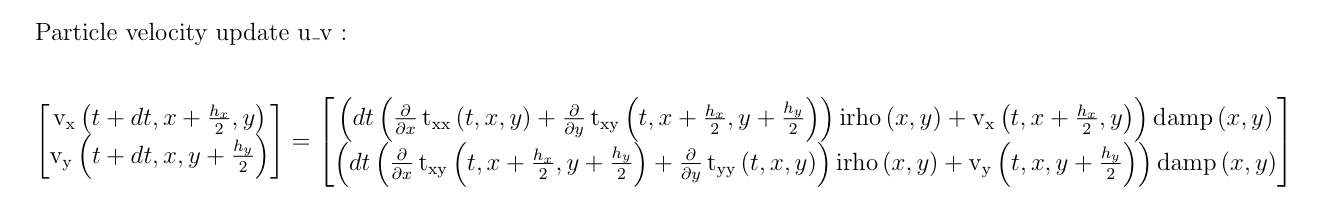}
\caption{Particle velocity update equation. The stress update
is omitted for readability (the equation does not fit into a single page).
    However, it can be found in a Devito
    \href{https://github.com/devitocodes/devito/blob/v4.2/examples/seismic/tutorials/06_elastic_varying_parameters.ipynb}{tutorial}
    on elastic modelling.}\label{PrettyElas}
\end{figure*}

The \texttt{SymPy} expressions created by these commands can be displayed
with \texttt{sympy.pprint} as shown in Figure~\ref{PrettyElas}. This
representation reflects perfectly the mathematics while maintaining
computational portability and efficiency through the \devito compiler. The
complete generated code for the elastic wave-equation with and without MPI
is made available at
\href{https://github.com/mloubout/SC20Paper/tree/master/codesamples}{CodeSample}
in \texttt{elastic-so12.c} and \texttt{elastic-so12-mpi.c}.

\subsection{2D example}\label{d-example}

\begin{figure*}
\centering
\includegraphics[width=1.000\hsize]{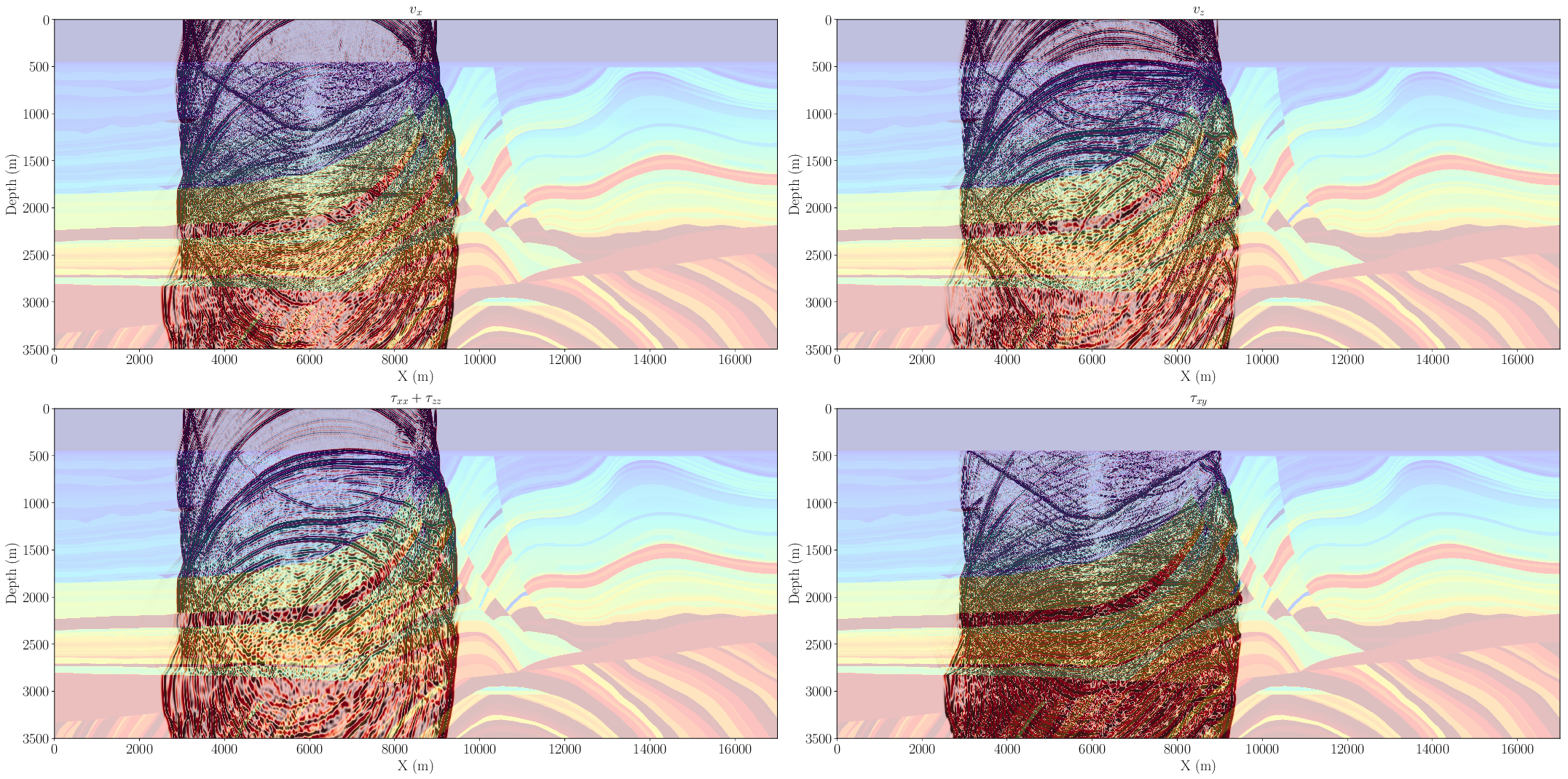}
\caption{Particle velocities and stress at time $t=3\text{s}$ for a
10m deep source and $x=5\text{km}$ in the
marmousi-ii model.}\label{ElasWf}
\end{figure*}

\begin{figure*}
\centering
\captionsetup[subfigure]{labelformat=empty}
\subfloat[]{\includegraphics[width=0.300\hsize]{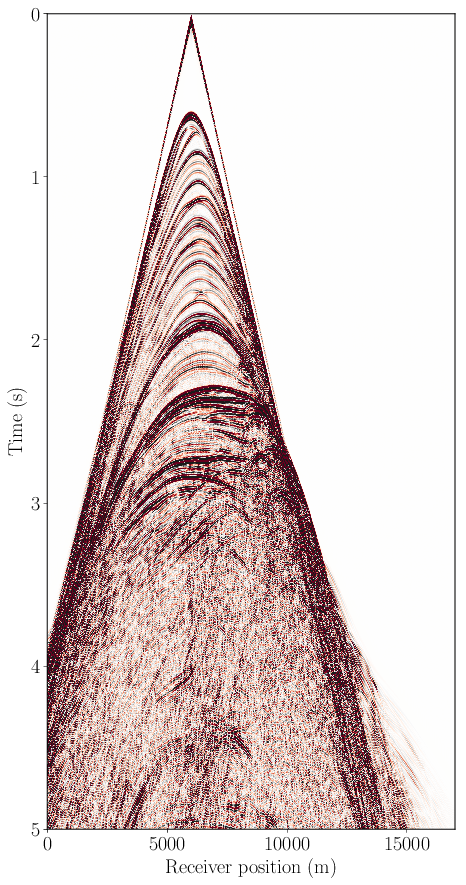}}
\subfloat[]{\includegraphics[width=0.300\hsize]{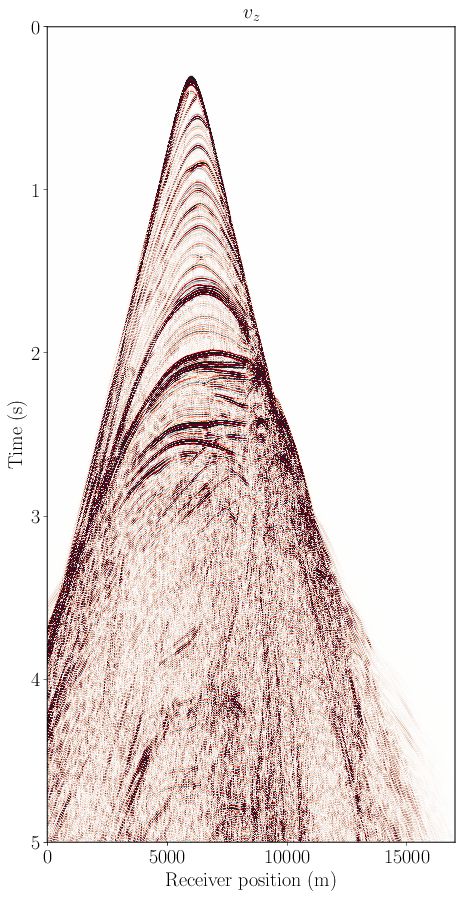}}
\subfloat[]{\includegraphics[width=0.300\hsize]{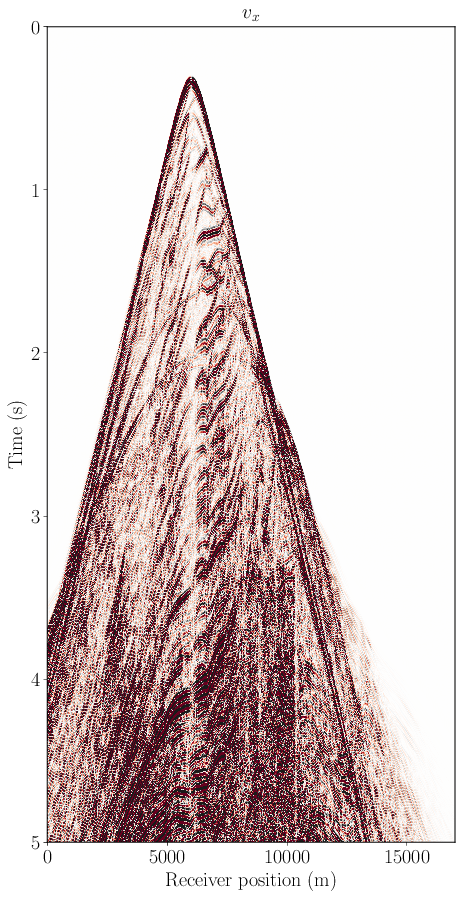}}
\caption{Seismic shot record for 5sec of modelling. \texttt{a} is the
pressure (trace of stress tensor) at the surface (5m depth), \texttt{b}
and \texttt{c} display, respectively, the vertical and horizontal
particle velocity at the ocean bottom (450m depth).}\label{ElasShot}
\end{figure*}

To demonstrate the efficacy of the elastic implementation outlined
above we utilized a broadly recognized 2D synthetic model, the elastic
Marmousi-ii\cite{versteeg927, marmouelas} model. The wavefields are
shown on Figure~\ref{ElasWf} and its corresponding elastic shot records
are displayed in Figure~\ref{ElasShot}. These two figures show that the
wavefield is, as expected, purely acoustic in the water layer
($\tau_{xy}=0$) and transitions correctly at the ocean bottom to
an elastic wavefield. We can also clearly see the shear wave-front
in the subsurface (at a depth of approximately 1km). Figures \ref{ElasWf}
and \ref{ElasShot} demonstrate that this high-level \devito implementation
of the elastic wave-equation is effective and accurate. Importantly, in
constructing this model within the \devito DSL framework, computational
technicalities such as the staggered grid analysis are abstracted away. We
note that the shot records displayed in Figure~\ref{ElasShot} match the
original data generated by the creator of this elastic model
(\cite{marmouelas}).

\subsection{3D proof of concept}\label{d-proof-of-concept}

Finally, three dimensional elastic data was modelled in the Cloud to
demonstrate the scalability of \devito to cluster-size problems. The model
used in this experiment mimics a reference model in geophysics known as
the SEAM model \cite{fehler2011seam}, a three dimensional extreme-scale
synthetic representation of a subsurface. The physical dimensions of the
model are \texttt{45kmx35kmx15km} discretized with a grid spacing of
\texttt{20mx20mx10m} leading to a computational grid of
\texttt{2250x1750x1500} grid points (5.9 billion grid points). One of the
main challenges of elastic modelling is the extreme memory cost owing to
the number of wavefields (a minimum of 21 fields in a three dimensional
propagator) that need to be stored:

\begin{itemize}
\itemsep1pt\parskip0pt\parsep0pt
\item
  Three particle velocities with two time steps (\texttt{v.forward} and
  \texttt{v})
\item
  Six stress with two time steps (\texttt{tau.forward} and \texttt{tau})
\item
  Three model parameters \texttt{lambda}, \texttt{mu} and \texttt{rho}
\end{itemize}

These 21 fields, for the 5.9 billion point grid defined above, lead to a
minimum memory requirement of 461Gb for modelling alone. For this experiment,
access was obtained for small HPC VMs (on Azure) called \texttt{Standard\_H16r}.
These VMs contain 16 core Intel Xeon E5 2667 v3 chips, with no hyperthreading,
and 32 nodes were used for a single source experiment
(i.e. a single wave-equation was solved). We
used a 12th order discretization in space that led to 2.8TFLOP/time-step being
computed by this model and the elastic wave was propagated for 16 seconds
(23000 time steps). Completion of this modelling run took 16 hours, converting to
1.1TFLOP/s. While these numbers may appear low, it should be noted that
the elastic kernel is extremely memory bound, while the TTI kernel is nearly
compute bound (see rooflines in \cite{louboutin2016ppf, devito-api, devito-compiler})
making it more computationally efficient, particularly in combination
with MPI. Future work will involve working on InfiniBand enabled and
true HPC VMs on Azure to achieve Cloud performance on par with that of state of
the art HPC clusters. Extrapolating from the performance obtained in this
experiment, and assuming a fairly standard setup of 5000 independent
source experiments, computing an elastic synthetic dataset would require 322
EFLOPs (23k time-steps x 2.8TFLOP/time-step x 5000
sources), or utilizing the full scalability and computing all sources in
parallel this becomes 5.5PFLOP/s.

\section{Performance comparison with other
codes}\label{performance-comparison-with-other-codes}

Earlier performance benchmarks mainly focused on roofline model analysis.
In this study, for completeness, the runtime of \devito is therefore
compared to that of the open source hand-coded propagator
\href{https://github.com/JanThorbecke/OpenSource.git}{fdelmodc}. This propagator,
described in \cite{thorbecke}, is a state of the art elastic kernel
(Equation~\ref{elas1}) and the comparisons presented here were carried out
in collaboration with its author. To ensure a fair
comparison, we ensured that the
physical and computational settings were identical. The settings were as
follows:

\begin{itemize}
\itemsep1pt\parskip0pt\parsep0pt
\item
  2001 by 1001 physical grid points.
\item
  200 grid points of dampening layer (absorbing layer \cite{cerjan}) on
  all four sides (total of 2401x1401 computational grid points).
\item
  10001 time steps.
\item
  Single point source, 2001 receivers.
\item
  Same compiler (GCC/ICC) to compile
  \href{https://github.com/JanThorbecke/OpenSource.git}{fdelmodc} and
  run Devito.
\item
  Intel(R) Xeon(R) CPU E3-1270 v6 @ 3.8GHz.
\item
  Single socket, four physical cores, four physical threads, thread
  pinning to cores and hyperthreading off.
\end{itemize}

The runtimes observed for this problem were essentially identical,
showing less than a one percent of difference. Such similar runtimes were
obtained with both the Intel and GNU compilers and the
experiment was performed with both fourth and sixth order discretizations.
Kernels were executed five times each and the
runtimes observed were consistently very similar. This comparison
illustrates the performance achieved with \devito is at least on par with
hand-coded propagators. Considering we do not take advantage of the Devito
compilers full capabilities in two dimensional cases, we are confident
that the code generated will be at least on par with the hand-coded version
for three dimensional problems and this comparison will be part of our future
work.

\section{Conclusions}\label{conclusions}

Transitioning from academic toy problems, such as the two-dimensional
acoustic wave-equation, to real-world applications can be challenging,
particularly if this transition is carried out as an afterthought.
Owing to the fundamental design principles of \devito such scaling,
however, becomes trivial. In this work we demonstrated
the high-level interface provided by \devito not only for simple scalar
equations but also for coupled PDEs. This interface allows, in a simple, concise
and consistent manner, the expression of all-kinds of non-trivial
differential operators. Next, and most
importantly, we demonstrated that the compiler enables large-scale modelling with
state-of-the art computational performance and programming paradigm. The
single-node performance is on par with state of the art hand-coded models, but
packaged with this performance comes
the flexibility of the symbolic interface and multi-node parallelism, which is
integrated in the compiler and interface in a accessible way. Finally,
we demonstrated that our abstractions provide the necessary portability
to enable both on-premise and Cloud based HPC.

\section{Code availability}

The code to reproduce the different examples presented in this work is available online in the following repositories:

\begin{itemize}
  \item The complete code for TTI imaging is available at \url{https://github.com/slimgroup/Azure2019/tree/v1.0} and includes 
  the TTI propagators, the Azure setup for shot parallelism and a documentation.
  \item The elastic modelling can be run with the elastic example available in Devito at
  \url{https://github.com/devitocodes/devito/blob/v4.2/examples/seismic/elastic/elastic_example.py} 
  and can be run with any size and spatial order. A standalone script to run the large 3D elastic modelling
  is also available at \url{https://github.com/mloubout/SC20Paper/tree/master/codesamples}
\end{itemize}

\bibliography{sc20_paper}

 \newcommand{\noop}[1]{}
\begin{thebibliography}{10}
\providecommand{\url}[1]{#1}
\csname url@samestyle\endcsname
\providecommand{\newblock}{\relax}
\providecommand{\bibinfo}[2]{#2}
\providecommand{\BIBentrySTDinterwordspacing}{\spaceskip=0pt\relax}
\providecommand{\BIBentryALTinterwordstretchfactor}{4}
\providecommand{\BIBentryALTinterwordspacing}{\spaceskip=\fontdimen2\font plus
\BIBentryALTinterwordstretchfactor\fontdimen3\font minus
  \fontdimen4\font\relax}
\providecommand{\BIBforeignlanguage}[2]{{%
\expandafter\ifx\csname l@#1\endcsname\relax
\typeout{** WARNING: IEEEtran.bst: No hyphenation pattern has been}%
\typeout{** loaded for the language `#1'. Using the pattern for}%
\typeout{** the default language instead.}%
\else
\language=\csname l@#1\endcsname
\fi
#2}}
\providecommand{\BIBdecl}{\relax}
\BIBdecl

\bibitem{lyu2020}
\BIBentryALTinterwordspacing
C.~Lyu, Y.~Capdeville, and L.~Zhao, ``Efficiency of the spectral element method
  with very high polynomial degree to solve the elastic wave equation,''
  \emph{GEOPHYSICS}, vol.~85, no.~1, pp. T33--T43, 2020. [Online]. Available:
  \url{https://doi.org/10.1190/geo2019-0087.1}
\BIBentrySTDinterwordspacing

\bibitem{fichtner2011}
A.~Fichtner, \emph{{Full Seismic Waveform Modelling and Inversion}}.\hskip 1em
  plus 0.5em minus 0.4em\relax Springer Verlag, 2010.

\bibitem{tarantola}
\BIBentryALTinterwordspacing
A.~Tarantola, ``Inversion of seismic reflection data in the acoustic
  approximation,'' \emph{GEOPHYSICS}, vol.~49, no.~8, p. 1259, 1984. [Online].
  Available: \url{+ http://dx.doi.org/10.1190/1.1441754}
\BIBentrySTDinterwordspacing

\bibitem{sirgue}
\BIBentryALTinterwordspacing
L.~Sirgue, O.~I.~Barkved, J.~P.~Van~Gestel, O.~J.~Askim, and J.~H.~Kommedal,
  ``3d waveform inversion on valhall wide-azimuth obc,'' 2009. [Online].
  Available:
  \url{https://www.earthdoc.org/content/papers/10.3997/2214-4609.201400395}
\BIBentrySTDinterwordspacing

\bibitem{ratcliffe2011}
\BIBentryALTinterwordspacing
A.~Ratcliffe, C.~Win, V.~Vinje, G.~Conroy, M.~Warner, A.~Umpleby, I.~Stekl,
  T.~Nangoo, and A.~Bertrand, ``Full waveform inversion: A north sea {OBC} case
  study,'' jan 2011. [Online]. Available:
  \url{https://doi.org/10.1190%2F1.3627688}
\BIBentrySTDinterwordspacing

\bibitem{petersson2013}
N.~Petersson and B.~Sj{\"o}green, \emph{SW4 v1.1 [software]}, Computational
  Infrastructure for Geodynamics, 2014.

\bibitem{osti_1468379}
L.~Preston, ``Computation of kernels for full waveform seismic inversion using
  parelasti.'' 8 2018.

\bibitem{osti_1561580}
------, ``Paraniso 1.0: 3-d full waveform seismic simulation in general
  anisotropic media.'' 9 2019.

\bibitem{osti_1561581}
------, ``Parelastifwi 1.0 user guide.'' 9 2019.

\bibitem{sava1}
\BIBentryALTinterwordspacing
D.~Köhn, O.~Hellwig, D.~De~Nil, and W.~Rabbel, ``Waveform inversion in
  triclinic anisotropic media—a resolution study,'' \emph{Geophysical Journal
  International}, vol. 201, no.~3, pp. 1642--1656, 04 2015. [Online].
  Available: \url{https://doi.org/10.1093/gji/ggv097}
\BIBentrySTDinterwordspacing

\bibitem{sava2}
\BIBentryALTinterwordspacing
B.~Zehner, O.~Hellwig, M.~Linke, I.~Görz, and S.~Buske, ``Rasterizing
  geological models for parallel finite difference simulation using seismic
  simulation as an example,'' \emph{Computers \& Geosciences}, vol.~86, pp. 83
  -- 91, 2016. [Online]. Available:
  \url{http://www.sciencedirect.com/science/article/pii/S009830041530073X}
\BIBentrySTDinterwordspacing

\bibitem{peter2011}
\BIBentryALTinterwordspacing
D.~Peter, D.~Komatitsch, Y.~Luo, R.~Martin, N.~Le~Goff, E.~Casarotti,
  P.~Le~Loher, F.~Magnoni, Q.~Liu, C.~Blitz, T.~Nissen-Meyer, P.~Basini, and
  J.~Tromp, ``Forward and adjoint simulations of seismic wave propagation on
  fully unstructured hexahedral meshes,'' \emph{Geophysical Journal
  International}, vol. 186, no.~2, pp. 721--739, 2011. [Online]. Available:
  \url{https://onlinelibrary.wiley.com/doi/abs/10.1111/j.1365-246X.2011.05044.x}
\BIBentrySTDinterwordspacing

\bibitem{krebsdg}
\BIBentryALTinterwordspacing
J.~Krebs, S.~Collis, N.~Downey, C.~Ober, J.~Overfelt, T.~Smith, B.~van
  Bloemen-Waanders, and J.~Young, ``Full wave inversion using a
  spectral-element discontinuous galerkin method,'' vol. 2014, no.~1, pp. 1--5,
  2014. [Online]. Available:
  \url{https://www.earthdoc.org/content/papers/10.3997/2214-4609.20140707}
\BIBentrySTDinterwordspacing

\bibitem{trinh2019}
P.-T. Trinh, R.~Brossier, L.~Métivier, L.~Tavard, and J.~Virieux, ``Efficient
  time-domain 3d elastic and viscoelastic full-waveform inversion using a
  spectral-element method on flexible cartesian-based mesh,''
  \emph{Geophysics}, vol.~84, no.~1, pp. R75--R97, January-February 2019.

\bibitem{virieuxmodelling}
\BIBentryALTinterwordspacing
J.~Virieux, S.~Operto, H.~Ben-Hadj-Ali, R.~Brossier, V.~Etienne, F.~Sourbier,
  L.~Giraud, and A.~Haidar, ``Seismic wave modeling for seismic imaging,''
  \emph{The Leading Edge}, vol.~28, no.~5, pp. 538--544, 2009. [Online].
  Available: \url{https://doi.org/10.1190/1.3124928}
\BIBentrySTDinterwordspacing

\bibitem{devito-api}
\BIBentryALTinterwordspacing
M.~Louboutin, M.~Lange, F.~Luporini, N.~Kukreja, P.~A. Witte, F.~J. Herrmann,
  P.~Velesko, and G.~J. Gorman, ``Devito (v3.1.0): an embedded domain-specific
  language for finite differences and geophysical exploration,''
  \emph{Geoscientific Model Development}, vol.~12, no.~3, pp. 1165--1187, 2019.
  [Online]. Available: \url{https://www.geosci-model-dev.net/12/1165/2019/}
\BIBentrySTDinterwordspacing

\bibitem{sympy}
\BIBentryALTinterwordspacing
A.~Meurer, C.~P. Smith, M.~Paprocki, O.~\v{C}ert\'{i}k, S.~B. Kirpichev,
  M.~Rocklin, A.~Kumar, S.~Ivanov, J.~K. Moore, S.~Singh, T.~Rathnayake,
  S.~Vig, B.~E. Granger, R.~P. Muller, F.~Bonazzi, H.~Gupta, S.~Vats,
  F.~Johansson, F.~Pedregosa, M.~J. Curry, A.~R. Terrel, v.~Rou\v{c}ka,
  A.~Saboo, I.~Fernando, S.~Kulal, R.~Cimrman, and A.~Scopatz, ``Sympy:
  symbolic computing in python,'' \emph{PeerJ Computer Science}, vol.~3, p.
  e103, Jan. 2017. [Online]. Available:
  \url{https://doi.org/10.7717/peerj-cs.103}
\BIBentrySTDinterwordspacing

\bibitem{fenics}
A.~Logg, K.-A. Mardal, G.~N. Wells \emph{et~al.}, \emph{Automated Solution of
  Differential Equations by the Finite Element Method}.\hskip 1em plus 0.5em
  minus 0.4em\relax Springer, 2012.

\bibitem{firedrake}
\BIBentryALTinterwordspacing
F.~Rathgeber, D.~A. Ham, L.~Mitchell, M.~Lange, F.~Luporini, A.~T.~T. McRae,
  G.~Bercea, G.~R. Markall, and P.~H.~J. Kelly, ``Firedrake: automating the
  finite element method by composing abstractions,'' \emph{CoRR}, vol.
  abs/1501.01809, 2015. [Online]. Available:
  \url{http://arxiv.org/abs/1501.01809}
\BIBentrySTDinterwordspacing

\bibitem{devito-compiler}
\BIBentryALTinterwordspacing
F.~{Luporini}, M.~{Lange}, M.~{Louboutin}, N.~{Kukreja}, J.~{H{\"u}ckelheim},
  C.~{Yount}, P.~{Witte}, P.~H.~J. {Kelly}, G.~J. {Gorman}, and F.~J.
  {Herrmann}, ``Architecture and performance of devito, a system for automated
  stencil computation,'' \emph{CoRR}, vol. abs/1807.03032, jul 2018. [Online].
  Available: \url{http://arxiv.org/abs/1807.03032}
\BIBentrySTDinterwordspacing

\bibitem{virieux}
\BIBentryALTinterwordspacing
J.~Virieux and S.~Operto, ``An overview of full-waveform inversion in
  exploration geophysics,'' \emph{GEOPHYSICS}, vol.~74, no.~5, pp. WCC1--WCC26,
  2009. [Online]. Available:
  \url{http://library.seg.org/doi/abs/10.1190/1.3238367}
\BIBentrySTDinterwordspacing

\bibitem{thomsen1986}
L.~{Thomsen}, ``Weak elastic anisotropy,'' \emph{Geophysics}, vol.~51, no.~10,
  pp. 1964--1966, october 1986.

\bibitem{zhang-tti}
\BIBentryALTinterwordspacing
Y.~Zhang, H.~Zhang, and G.~Zhang, ``A stable tti reverse time migration and its
  implementation,'' \emph{GEOPHYSICS}, vol.~76, no.~3, pp. WA3--WA11, 2011.
  [Online]. Available: \url{https://doi.org/10.1190/1.3554411}
\BIBentrySTDinterwordspacing

\bibitem{duveneck}
\BIBentryALTinterwordspacing
E.~Duveneck and P.~M. Bakker, ``Stable p-wave modeling for reverse-time
  migration in tilted ti media,'' \emph{GEOPHYSICS}, vol.~76, no.~2, pp.
  S65--S75, 2011. [Online]. Available: \url{https://doi.org/10.1190/1.3533964}
\BIBentrySTDinterwordspacing

\bibitem{louboutin2018segeow}
\BIBentryALTinterwordspacing
M.~Louboutin, P.~A. Witte, and F.~J. Herrmann, ``Effects of wrong adjoints for
  rtm in tti media,'' in \emph{SEG Technical Program Expanded Abstracts}, 10
  2018, pp. 331--335, (SEG, Anaheim). [Online]. Available:
  \url{https://slim.gatech.edu/Publications/Public/Conferences/SEG/2018/louboutin2018SEGeow/louboutin2018SEGeow.html}
\BIBentrySTDinterwordspacing

\bibitem{patterson}
D.~A. Patterson and J.~L. Hennessy, \emph{Computer Organization and Design: The
  Hardware/Software Interface}, 3rd~ed.\hskip 1em plus 0.5em minus 0.4em\relax
  San Francisco, CA, USA: Morgan Kaufmann Publishers Inc., 2007.

\bibitem{louboutin2016ppf}
M.~Louboutin, M.~Lange, F.~J. Herrmann, N.~Kukreja, and G.~Gorman,
  ``Performance prediction of finite-difference solvers for different computer
  architectures,'' \emph{Computers \& Geosciences}, vol. 105, pp. 148--157, 08
  2017.

\bibitem{witte2018alf}
\BIBentryALTinterwordspacing
P.~A. Witte, M.~Louboutin, N.~Kukreja, F.~Luporini, M.~Lange, G.~J. Gorman, and
  F.~J. Herrmann, ``A large-scale framework for symbolic implementations of
  seismic inversion algorithms in julia,'' \emph{Geophysics}, vol.~84, no.~3,
  pp. F57--F71, 03 2019, (Geophysics). [Online]. Available:
  \url{https://slim.gatech.edu/Publications/Public/Journals/Geophysics/2019/witte2018alf/witte2018alf.pdf}
\BIBentrySTDinterwordspacing

\bibitem{alkhalifah2000}
T.~{Alkhalifah}, ``An acoustic wave equation for anisotropic media,''
  \emph{Geophysics}, vol.~65, no.~4, pp. 1239--1250, 2000.

\bibitem{baysal1983}
E.~{Baysal}, D.~D. {Kosloff}, , and J.~W.~C. {Sherwood}, ``Reverse time
  migration,'' \emph{Geophysics}, vol.~48, no.~11, pp. 1514--1524, november
  1983.

\bibitem{bubetti2012}
\BIBentryALTinterwordspacing
K.~P. Bube, T.~Nemeth, J.~P. Stefani, R.~Ergas, W.~Liu, K.~T. Nihei, and
  L.~Zhang, ``On the instability in second-order systems for acoustic vti and
  tti media,'' \emph{GEOPHYSICS}, vol.~77, no.~5, pp. T171--T186, 2012.
  [Online]. Available: \url{https://doi.org/10.1190/geo2011-0250.1}
\BIBentrySTDinterwordspacing

\bibitem{bubetti2014}
\BIBentryALTinterwordspacing
K.~P. Bube*, R.~Ergas, and T.~Nemeth, \emph{Stability and energy conservation
  for second-order acoustic systems for VTI andTTI media with positive shear
  wavespeeds}.\hskip 1em plus 0.5em minus 0.4em\relax SEG, 2014, pp.
  3439--3443. [Online]. Available:
  \url{https://library.seg.org/doi/abs/10.1190/segam2014-0986.1}
\BIBentrySTDinterwordspacing

\bibitem{bubesatti2016}
\BIBentryALTinterwordspacing
K.~Bube, J.~Washbourne, R.~Ergas, and T.~Nemeth, \emph{Self-adjoint,
  energy-conserving second-order pseudoacoustic systems for VTI and TTI media
  for reverse time migration and full-waveform inversion}.\hskip 1em plus 0.5em
  minus 0.4em\relax SEG, 2016, pp. 1110--1114. [Online]. Available: \url{e}
\BIBentrySTDinterwordspacing

\bibitem{chu2011}
\BIBentryALTinterwordspacing
C.~Chu, B.~K. Macy, and P.~D. Anno, ``Approximation of pure acoustic seismic
  wave propagation in tti media,'' \emph{GEOPHYSICS}, vol.~76, no.~5, pp.
  WB97--WB107, 2011. [Online]. Available:
  \url{https://doi.org/10.1190/geo2011-0092.1}
\BIBentrySTDinterwordspacing

\bibitem{fletcher}
\BIBentryALTinterwordspacing
R.~P. Fletcher, X.~Du, and P.~J. Fowler, ``Reverse time migration in tilted
  transversely isotropic (tti) media,'' \emph{GEOPHYSICS}, vol.~74, no.~6, pp.
  WCA179--WCA187, 2009. [Online]. Available:
  \url{https://doi.org/10.1190/1.3269902}
\BIBentrySTDinterwordspacing

\bibitem{fowlertti2010}
\BIBentryALTinterwordspacing
P.~J. Fowler, X.~Du, and R.~P. Fletcher, ``Coupled equations for reverse time
  migration in transversely isotropic media,'' \emph{GEOPHYSICS}, vol.~75,
  no.~1, pp. S11--S22, 2010. [Online]. Available:
  \url{https://doi.org/10.1190/1.3294572}
\BIBentrySTDinterwordspacing

\bibitem{whitmore1983}
N.~D. {Whitmore}, ``Iterative depth migration by backward time propagation,''
  \emph{1983 SEG Annual Meeting, Expanded Abstracts}, 1983.

\bibitem{witte2016segpve}
\BIBentryALTinterwordspacing
P.~A. Witte, C.~C. Stolk, and F.~J. Herrmann, ``Phase velocity error minimizing
  scheme for the anisotropic pure p-wave equation,'' in \emph{SEG Technical
  Program Expanded Abstracts}, 10 2016, pp. 452--457, (SEG, Dallas). [Online].
  Available:
  \url{https://slim.gatech.edu/Publications/Public/Conferences/SEG/2016/witte2016SEGpve/witte2016SEGpve.html}
\BIBentrySTDinterwordspacing

\bibitem{xu2014}
S.~{Xu} and H.~{Zhou}, ``Accurate simulations of pure quasi-p-waves in complex
  anisotropic media,'' \emph{Geophysics}, vol.~79, no.~6, pp. 341--348,
  november-december 2014.

\bibitem{zhang2005}
L.~{Zhang}, J.~W. {Rector III}, and H.~G. Micheal, ``Finite-difference
  modelling of wave propagation in acoustic tilted ti media,''
  \emph{Geophysical Prospecting}, vol.~53, pp. 843--852, 2005.

\bibitem{zhang2011}
Y.~{Zhang}, H.~{Zhang}, and G.~{Zhang}, ``A stable tti reverse time migration
  and its implementation,'' \emph{Geophysics}, vol.~76, no.~3, pp. WA3--WA11,
  may-june 2011.

\bibitem{zhan2013}
G.~{Zhan}, R.~C. {Pestana}, and P.~L. {Stoffa}, ``An efficient hybrid
  pdeudospectral/finite-difference scheme for solving the tti pure p-wave
  equation,'' \emph{Journal of Geophyics and Engineering}, vol.~10, 2013.

\bibitem{hooke}
\BIBentryALTinterwordspacing
R.~Hooke, D.~Papin, S.~Sturmy, and J.~Young, \emph{Lectures de potentia
  restitutiva}.\hskip 1em plus 0.5em minus 0.4em\relax London : Printed for
  John Martyn ..., 1678. [Online]. Available:
  \url{http://lib.ugent.be/catalog/rug01:001559640}
\BIBentrySTDinterwordspacing

\bibitem{louboutin2020THmfi}
\BIBentryALTinterwordspacing
M.~Louboutin, ``Modeling for inversion in exploration geophysics,'' phd,
  Georgia Institute of Technology, Atlanta, 03 2020, (PhD). [Online].
  Available:
  \url{https://slim.gatech.edu/Publications/Public/Thesis/2020/louboutin2020THmfi/louboutin2020THmfi.pdf}
\BIBentrySTDinterwordspacing

\bibitem{witte2019TPDedas}
P.~A. {Witte}, M.~{Louboutin}, H.~{Modzelewski}, C.~{Jones}, J.~{Selvage}, and
  F.~J. {Herrmann}, ``An event-driven approach to serverless seismic imaging in
  the cloud,'' 2020.

\bibitem{witte2019SEGedw}
\BIBentryALTinterwordspacing
P.~A. Witte, M.~Louboutin, H.~Modzelewski, C.~Jones, J.~Selvage, and F.~J.
  Herrmann, ``Event-driven workflows for large-scale seismic imaging in the
  cloud,'' in \emph{SEG Technical Program Expanded Abstracts}, 09 2019, pp.
  3984--3988, (SEG, San Antonio). [Online]. Available:
  \url{https://slim.gatech.edu/Publications/Public/Conferences/SEG/2019/witte2019SEGedw/witte2019SEGedw.html}
\BIBentrySTDinterwordspacing

\bibitem{herrmann2019EAGEHPCaii}
F.~J. Herrmann, C.~Jones, G.~Gorman, J.~H{\"u}ckelheim, K.~Lensink, P.~Kelly,
  N.~Kukreja, H.~Modzelewski, M.~Lange, M.~Louboutin, F.~Luporini, J.~Selvages,
  and P.~A. Witte, ``Accelerating ideation \& innovation cheaply in the cloud
  the power of abstraction, collaboration \& reproducibility,'' in \emph{4th
  EAGE Workshop on High-performance Computing}, 10 2019, (EAGE HPC Workshop,
  Dubai).

\bibitem{witte2019RHPCssi}
\BIBentryALTinterwordspacing
P.~A. Witte, M.~Louboutin, C.~Jones, and F.~J. Herrmann, ``Serverless seismic
  imaging in the cloud,'' 2019, submitted to Rice Oil and Gas High Performance
  Computing Conference 2020 on November 27, 2019. [Online]. Available:
  \url{https://slim.gatech.edu/Publications/Public/Submitted/2019/witte2019RHPCssi/witte2019RHPCssi.html}
\BIBentrySTDinterwordspacing

\bibitem{dask}
M.~Rocklin, ``Dask: Parallel computation with blocked algorithms and task
  scheduling,'' in \emph{Proceedings of the 14th Python in Science Conference},
  K.~Huff and J.~Bergstra, Eds., 2015, pp. 130 -- 136.

\bibitem{ufl}
M.~S. Aln{\ae}s, A.~Logg, K.~B. {\O}lgaard, M.~E. Rognes, and G.~N. Wells,
  ``{U}nified {F}orm {L}anguage: a domain-specific language for weak
  formulations of partial differential equations,'' \emph{ACM Transactions on
  Mathematical Software (TOMS)}, vol.~40, no.~2, p.~9, 2014.

\bibitem{versteeg927}
\BIBentryALTinterwordspacing
R.~Versteeg, ``The marmousi experience; velocity model determination on a
  synthetic complex data set,'' \emph{The Leading Edge}, vol.~13, no.~9, pp.
  927--936, 1994. [Online]. Available:
  \url{http://tle.geoscienceworld.org/content/13/9/927}
\BIBentrySTDinterwordspacing

\bibitem{marmouelas}
\BIBentryALTinterwordspacing
G.~S. Martin, R.~Wiley, and K.~J. Marfurt, ``Marmousi2: An elastic upgrade for
  marmousi,'' \emph{The Leading Edge}, vol.~25, no.~2, pp. 156--166, 2006.
  [Online]. Available: \url{https://doi.org/10.1190/1.2172306}
\BIBentrySTDinterwordspacing

\bibitem{fehler2011seam}
M.~Fehler and P.~J. Keliher, \emph{SEAM Phase 1: Challenges of subsalt imaging
  in tertiary basins, with emphasis on deepwater Gulf of Mexico}.\hskip 1em
  plus 0.5em minus 0.4em\relax Society of Exploration Geophysicists, 2011.

\bibitem{thorbecke}
\BIBentryALTinterwordspacing
J.~W. Thorbecke and D.~Draganov, ``Finite-difference modeling experiments for
  seismic interferometry,'' \emph{GEOPHYSICS}, vol.~76, no.~6, pp. H1--H18,
  2011. [Online]. Available: \url{https://doi.org/10.1190/geo2010-0039.1}
\BIBentrySTDinterwordspacing

\bibitem{cerjan}
\BIBentryALTinterwordspacing
C.~Cerjan, D.~Kosloff, R.~Kosloff, and M.~Reshef, ``A nonreflecting boundary
  condition for discrete acoustic and elastic wave equations,''
  \emph{GEOPHYSICS}, vol.~50, no.~4, pp. 705--708, 1985. [Online]. Available:
  \url{https://doi.org/10.1190/1.1441945}
\BIBentrySTDinterwordspacing

\end{thebibliography}

\end{document}